\documentclass[12pt]{article}
\begin{document}
      \sloppy

\def\AFOUR{%
\setlength{\textheight}{9.0in}%
\setlength{\textwidth}{5.75in}%
\setlength{\topmargin}{-0.375in}%
\hoffset=-.5in%
\renewcommand{\baselinestretch}{1.17}%
\setlength{\parskip}{6pt plus 2pt}%
}
\AFOUR
\def\car{\mathop{\square}}
\def\carre#1#2{\raise 2pt\hbox{$\scriptstyle #1$}\car_{#2}}

\parindent=0pt
\makeatletter
\def\section{\@startsection {section}{1}{\z@}{-3.5ex plus -1ex minus
   -.2ex}{2.3ex plus .2ex}{\large\bf}}
\def\subsection{\@startsection{subsection}{2}{\z@}{-3.25ex plus -1ex
minus
   -.2ex}{1.5ex plus .2ex}{\normalsize\bf}}
\makeatother
\makeatletter
\@addtoreset{equation}{section}
\renewcommand{\theequation}{\thesection.\arabic{equation}}
\makeatother

\renewcommand{\a}{\alpha}
\renewcommand{\b}{\beta}
\newcommand{\g}{\gamma}           \newcommand{\G}{\Gamma}
\renewcommand{\d}{\delta}         \newcommand{\D}{\Delta}
\newcommand{\e}{\varepsilon}
\newcommand{\la}{\lambda}        \newcommand{\LA}{\Lambda}
\newcommand{\m}{\mu}
\newcommand{\A}{\widehat{A}^{\star a}_{\mu}}
\newcommand{\Ar}{\widehat{A}^{\star a}_{\rho}}
\newcommand{\n}{\nu}
\newcommand{\om}{\omega}         \newcommand{\OM}{\Omega}
\newcommand{\p}{\psi}             \newcommand{\PS}{\Psi}
\renewcommand{\r}{\rho}
\newcommand{\s}{\sigma}           \renewcommand{\S}{\Sigma}
\newcommand{\vf}{{\varphi}}
\newcommand{\y}{{\upsilon}}       \newcommand{\Y}{{\Upsilon}}
\newcommand{\z}{\zeta}

\renewcommand{\AA}{{\cal A}}
\newcommand{\BB}{{\cal B}}
\newcommand{\CC}{{\cal C}}
\newcommand{\DD}{{\cal D}}
\newcommand{\EE}{{\cal E}}
\newcommand{\FF}{{\cal F}}
\newcommand{\GG}{{\cal G}}
\newcommand{\HH}{{\cal H}}
\newcommand{\II}{{\cal I}}
\newcommand{\JJ}{{\cal J}}
\newcommand{\KK}{{\cal K}}
\newcommand{\LL}{{\cal L}}
\newcommand{\MM}{{\cal M}}
\newcommand{\NN}{{\cal N}}
\newcommand{\OO}{{\cal O}}
\newcommand{\PP}{{\cal P}}
\newcommand{\QQ}{{\cal Q}}
\renewcommand{\SS}{{\cal S}}
\newcommand{\RR}{{\cal R}}
\newcommand{\TT}{{\cal T}}
\newcommand{\UU}{{\cal U}}
\newcommand{\VV}{{\cal V}}
\newcommand{\WW}{{\cal W}}
\newcommand{\XX}{{\cal X}}
\newcommand{\YY}{{\cal Y}}
\newcommand{\ZZ}{{\cal Z}}

\newcommand{\ch}{\widehat{C}}
\newcommand{\gh}{\widehat{\gamma}}
\newcommand{\W}{W_{i}}
\newcommand{\na}{\nabla}
\newcommand{\xint}{\dint d^4x\;}
\newcommand{\sla}{\raise.15ex\hbox{$/$}\kern -.57em}
\newcommand{\Sla}{\raise.15ex\hbox{$/$}\kern -.70em}
\def\h{\hbar}
\def\Lp{\displaystyle{\biggl(}}
\def\Rp{\displaystyle{\biggr)}}
\def\LP{\displaystyle{\Biggl(}}
\def\RP{\displaystyle{\Biggr)}}
\newcommand{\lp}{\left(}\newcommand{\rp}{\right)}
\newcommand{\lc}{\left[}\newcommand{\rc}{\right]}
\newcommand{\lac}{\left\{}\newcommand{\rac}{\right\}}
\newcommand{\identity}{\bf 1\hspace{-0.4em}1}
\newcommand{\complex}{{\kern .1em {\raise .47ex
\hbox {$\scriptscriptstyle |$}}
      \kern -.4em {\rm C}}}
\newcommand{\real}{{{\rm I} \kern -.19em {\rm R}}}
\newcommand{\rational}{{\kern .1em {\raise .47ex
\hbox{$\scripscriptstyle |$}}
      \kern -.35em {\rm Q}}}
\renewcommand{\natural}{{\vrule height 1.6ex width
.05em depth 0ex \kern -.35em {\rm N}}}
\newcommand{\tint}{\int d^4 \! x \, }
\newcommand{\intg}{\int d^D \! x \, }
\newcommand{\intm}{\int_\MM}
\newcommand{\tr}{{\rm {Tr} \,}}
\newcommand{\half}{\dfrac{1}{2}}
\newcommand{\f}{\frac}
\newcommand{\pa}{\partial}
\newcommand{\pad}[2]{{\frac{\partial #1}{\partial #2}}}
\newcommand{\fud}[2]{{\frac{\delta #1}{\delta #2}}}
\newcommand{\dpad}[2]{{\displaystyle{\frac{\partial #1}{\partial
#2}}}}
\newcommand{\dfud}[2]{{\displaystyle{\frac{\delta #1}{\delta #2}}}}
\newcommand{\dfrac}[2]{{\displaystyle{\frac{#1}{#2}}}}
\newcommand{\dsum}[2]{\displaystyle{\sum_{#1}^{#2}}}
\newcommand{\dint}{\displaystyle{\int}}
\newcommand{\eg}{{\em e.g.,\ }}
\newcommand{\Eg}{{\em E.g.,\ }}
\newcommand{\ie}{{{\em i.e.},\ }}
\newcommand{\Ie}{{\em I.e.,\ }}
\newcommand{\nb}{\noindent{\bf N.B.}\ }
\newcommand{\etal}{{\em et al.}}
\newcommand{\etc}{{\em etc.\ }}
\newcommand{\via}{{\em via\ }}
\newcommand{\cf}{{\em cf.\ }}
\newcommand{\twiddle}{\lower.9ex\rlap{$\kern -.1em\scriptstyle\sim$}}
\newcommand{\qed}{\vrule height 1.2ex width 0.5em}
\newcommand{\grad}{\nabla}
\newcommand{\bra}[1]{\left\langle {#1}\right|}
\newcommand{\ket}[1]{\left| {#1}\right\rangle}
\newcommand{\vev}[1]{\left\langle {#1}\right\rangle}

\newcommand{\equ}[1]{(\ref{#1})}
\newcommand{\be}{\begin{equation}}
\newcommand{\ee}{\end{equation}}
\newcommand{\eqn}[1]{\label{#1}\end{equation}}
\newcommand{\eea}{\end{eqnarray}}
\newcommand{\bea}{\begin{eqnarray}}
\newcommand{\eqan}[1]{\label{#1}\end{eqnarray}}
\newcommand{\ba}{\begin{array}}
\newcommand{\ea}{\end{array}}
\newcommand{\eqac}{\begin{equation}\begin{array}{rcl}}
\newcommand{\eqacn}[1]{\end{array}\label{#1}\end{equation}}
\newcommand{\qq}{&\qquad &}
\renewcommand{\=}{&=&} 
\newcommand{\cb}{{\bar c}}
\newcommand{\mn}{{\m\n}}
\newcommand{\pic}{$\spadesuit\spadesuit$}
\newcommand{\?}{{\bf ???}}
\newcommand{\Tr }{\mbox{Tr}\ }
\newcommand{\adot}{{\dot\alpha}}
\newcommand{\bdot}{{\dot\beta}}
\newcommand{\gdot}{{\dot\gamma}}

\global\parskip=4pt
\titlepage  \noindent
{
   \noindent

\hfill GEF-TH-01/2008 


\vspace{2cm}

\noindent
{\bf
{\large Symanzik's Method Applied To The Fractional Quantum Hall Edge 
States
}}

\vspace{.5cm}
\hrule

\vspace{1cm}

\noindent
{\bf 
Alberto Blasi, Dario Ferraro, Nicola Maggiore, Nicodemo
Magnoli}

\noindent
{\footnotesize {\it
 Dipartimento di Fisica -- Universit\`a di Genova --
via Dodecaneso 33 -- I-16146 Genova -- Italy and INFN, Sezione di
Genova 
} }

{\bf Maura Sassetti}

\noindent
{\footnotesize {\it
 Dipartimento di Fisica -- Universit\`a di Genova --
via Dodecaneso 33 -- I-16146 Genova -- Italy and LAMIA-INFM-CNR.
} }

\vspace{1cm}
\noindent
{\tt Abstract~:}

In this paper we consider an abelian Chern-Simons theory with plane 
boundary and we show, following Symankiz's quite general  approach, 
how the known results for edge states in the Laughlin series can be 
derived in a systematic way by the separability condition. Moreover
we show that the conserved boundary currents find a natural and 
explicit interpretation in terms of the continuity equation and the 
Tomonaga-Luttinger commutation relation for electronic density
is recovered.

\vfill\noindent
{\footnotesize {\tt Keywords:}
Chern-Simons Theories,
Topological Field Theories,
Fractional Quantum Hall Effect
\\
{\tt PACS Nos:} 
03.70.+k Theory of Quantized Fields,
11.15.-q Gauge Field Theories,
73.43.-f Quantum Hall Effects
}
\newpage
\begin{small}
\end{small}

\setcounter{footnote}{0}


\section{Introduction}
The condensed matter theoreticians have adopted, quite a few years
ago, the three dimensional abelian Chern-Simons gauge model as one,
and maybe amongst the most important, ingredient to describe the low energy
physics of the Fractional Quantum Hall Effect (FQHE) 
\cite{Wen90,Wen91,Prange}. 
As is well known one is dealing with a topological theory that 
describes, by adding a coupling with an external electromagnetic
potential
and  with a current of gapped quasiparticles, the quantized Hall
conductance, 
the charge and the statistics of these quasiparticles. 
These results are insensitive to the details of 
the particular setup of the experimental apparatus and to the quantum 
fluctuations.\newline
A popular choice is to
add a boundary where the gapped quasiparticles  become
gapless (edge states) \cite{Wen91,Halperin,Cappelli95}. 
At this point one is faced with the problem of
choosing the boundary conditions for the gauge field and of computing
the action for the field theory which lives on the boundary.\newline
Many proposals have appeared in the Literature 
\cite{Wen91,Kane95,Fradkin99}; all of them achieve the goal of describing 
the wanted properties of the boundary field theory, but the ways to 
derive them are often based on 
{\it ad hoc} assumptions (for instance, the explicit 
dependence of the results on a particular choice of the 
gauge fixing term and/or of the boundary conditions).\newline
Here we reconsider the problem with a different approach, which stems
from the inclusion of a boundary 
which meets a precise physical requirement.
We would like to emphasize, that we will not find any new result;
the novelty is the method itself which fits in a rigorous,
perturbative,
quantum field theory treatment of the problem.
Now the inclusion of a boundary in a Minkowskian or Euclidean field
theory is not an easy task if one wishes to preserve locality and
power counting, the most basic ingredients of any perturbative
quantum field theory model.\newline
Many years ago K. Symanzik \cite{Symanzik81} proposed a solution: his
idea was to add to the classical bulk action a local boundary term
which modifies the propagator of the field in such a way that nothing
propagates from one side of the boundary to the other. He called this
property ``separability'' and showed that it requires a well identified
class of boundary conditions to be realized.\newline
Symanzik himself applied the method to compute the Casimir effect for
two parallel plates; later on there have been other applications
which include the non abelian Chern-Simons model where it is proven
that on the boundary there is a set of chiral currents obeying a
Kac-Moody algebra with central charge \cite{Blasi91}.
In this paper we apply Symanzik's method to the case of an abelian
Chern-Simons gauge model with a plane boundary; in order to make the
analysis as simple as possible, without losing rigor, we shall adopt
the gauge fixing of reference \cite{Piguet91}.
With this choice one can avoid all technical problems related to the
BRS transformations and to the presence of ghost fields. The gauge
symmetry, including the boundary contribution which is deduced from
the separability condition, is encoded in a local Ward identity; from
it we are able to compute, in a unique way, the propagators on the
boundary and therefore we identify the action which describes the
theory restricted to the boundary.
This action is written in terms of a scalar field which obeys a
chirality condition.\newline
Let us emphasize that this result is obtained without any 
{\it ad hoc} 
hypothesis concerning the gauge fixing term and/or the boundary conditions; 
identical conclusions can be reached by using, for instance, the Landau 
gauge at the price of some algebraic complications, and the boundary 
values of the field are given  by  the separability condition.\newline
The paper is so organized: in Section 2 we describe the bulk model 
just to fix the notation. The boundary is introduced in Section 3 
where we also identify the propagators obeying the separability
condition. 
The corresponding  effective bosonic action is given in Section 4. 
In  Section 5 we show that the chiral  boundary currents obey a
Kac-Moody 
algebra with central charge. Some conclusive remarks are collected in
the 
final Section, while in the Appendix, the propagator for the two 
dimensional effective bosonic action is derived.

\section{The Model}

\subsection{The Action}

Let us consider the abelian Chern-Simons theory
\be
S_{cs}=i\ \int d^{3}x\ \left (\frac{k}{2}\epsilon^{\mu\nu\rho}
A_{\mu}\partial_{\nu}A_{\rho}
+j^{\mu}A_{\mu}
\right ) ,
\label{action}
\ee
where $k$ is a coupling constant and $A_{\mu}(x)$ is a gauge field.
The
metric of the space is chosen to be flat euclidean, the infinitesimal
distance being
\be
(ds)^{2}=(dx_{0})^{2}+(dx_{1})^{2}+(dx_{2})^{2}\ .
\ee
In (\ref{action}) we introduce a coupling with a generic classical 
external current
$j^{\mu}$, whose normalization fixes the value of $k$.

The model defined by \equ{action} is consistent with
the quantization of the conductance for the Laughlin's sequence  
\cite{Laughlin83} 
of the FQHE, if
\be
k=\f{1}{2\pi \nu}=\f{2m+1}{2\pi}
\label{nu}
\ee
with $m\in \textbf{N}$ \cite{Wen90}. Here,
\be
\nu=\f{1}{2m+1}
\ee
represents the filling factor of the Quantum Hall fluid for the
Laughlin
sequence.
The relation between $k$ and $\nu$ is necessary to describe the
right properties of charge and statistic
of the quasiparticles that constitute the Hall fluid.\newline
For the aim of this paper, the presence of the classical current 
$j^{\mu}$ is irrelevant, and therefore it will be omitted in what 
follows, just keeping in mind that the coupling constant $k$ cannot 
be reabsorbed by a redefinition of the gauge fields $A_{\mu}(x)$.
In light-cone coordinates 
\begin{eqnarray}
u &=& x_{2}\nonumber \\
z &=& \frac{1}{\sqrt{2}}({x_{1}-ivx_{0}})  \label{lc}\\
\bar{z} &=&\frac{1}{\sqrt{2}}({x_{1} +ivx_{0}})\ ,  \nonumber
\end{eqnarray}
where $v$ is a non-relativistic velocity, 
the Chern-Simons action reads
\be
S_{cs}=i k\int dudzd\bar{z}\
\left(
\bar{A}\partial_{u}A + 
A_{u}\partial\bar{A} - A_{u}\bar\partial A
\right)\ ,
\ee
where
\begin{eqnarray}
A_{u}&=& A_{2} \nonumber \\
A&=& \frac{1}{\sqrt{2}}(A_{1}+\frac{i}{v}A_{0}) \label{lcfields}\\
\bar{A}&=& \frac{1}{\sqrt{2}}(A_{1}-\frac{i}{v}A_{0})\ .
\nonumber
\end{eqnarray}
We need now to introduce a gauge fixing term in the action, and we 
choose an  axial gauge condition
\be
A_{u}=0.
\label{gc}\ee
The ghost fields decouple completely from the
gauge field and so can be eliminated. Therefore the
gauge fixing term of the action is
\be
S_{gf}=i\int dudzd\bar{z}\ b A_{u}
\ee
where $b$ is a lagrangian multiplier.\newline
The complete action for the theory that we consider 
is therefore 
\be
S=S_{cs}+S_{gf}.
\ee

\subsection{Symmetries And Constraints}

To each object appearing in the action $S$, a canonical mass 
dimension and an ``helicity'' quantum number is assigned, as shown in 
Table 1~:
$$
\stackrel{Table\;1\;:\;Quantum\;Numbers}{
\begin{tabular}{|c|c|c|c|c|c|c|c|c|c|c|}
\hline 
& $ $ & $ $ & $ $ & $ $ & $ $ & $ $ & $ $ & $ $ & $ $ & $ $ \\
& $A_{u}$&$A$ & $\bar{A} $ & $b$ & $\partial_{u}$ & $\partial$ &
$\bar{\partial}$ & $u$ &$z$ &$\bar{z}$\\ \hline
$\dim $ & $1$&$1$ & $1$ & $2$ & $1$ &$1$& $1$ &$-1$&$-1$&$-1$ \\
\hline
$\mathrm{hel}$ & $0$&$1$ & $-1$ & $0$ & $0$ &$1$&$-1$ & $0$ & $-1$
&$1$ \\ \hline 
\end{tabular}
}  $$
The action $S$ is the most general one which obeys the following
constraints~:
\begin{enumerate}
\item $S$
is a dimensionless, local, integrated functional of 
a polynomial lagrangian 
density, with helicity 
zero.
\item $S$ is invariant under the local, infinitesimal gauge 
transformation
\begin{eqnarray}
    \delta A_{\mu} &=& \partial_{\mu}\theta \label{gauge} \\
    \delta b &=& 0\ , \nonumber 
    \end{eqnarray}
where $\theta$ is the local gauge 
parameter.
\item $S$ is invariant under the discrete symmetry involving at
the same time coordinates and fields
\begin{eqnarray}
	z &\leftrightarrow& \bar{z} \nonumber \\
	u &\rightarrow& -u \nonumber \\
	A &\leftrightarrow& \bar{A} \label{inversion}\\
	A_{u} &\rightarrow& - A_{u} \nonumber \\
	b &\rightarrow& -b \nonumber
\end{eqnarray}
\end{enumerate}

\subsection{The Generating Functional $Z[J]$}

Starting from the action $S$, we can define, in three dimensional 
euclidean spacetime, the generating functional
of the Green functions
\be
Z[J_{\chi}]=\int 
{\cal D}\chi
\exp \left [-\left ( S+\int du dz d\bar{z}\
\sum_{\chi}J_{\chi}\chi\right)\right ]
\label{Z}
\ee
with quantum sources for the gauge fields~: 
\be
\vev{\chi(X)}=\left.
\f{\delta Z}{\delta J_{\chi}(X)}
\right |_{J=0}
\ee
where $J_{\chi} = J_u , \bar{J}, J, J_b$, $\chi
=A_{u},A,\bar{A},b$, and $(X)\equiv (u,z,\bar{z})$).\newline
A generic $N$-point Green function is defined by 
\be
\vev{\chi_{1}(X_{1})\cdots\chi_{N}(X_{N})}=\left .
\f{\delta^{N}Z}{\delta
J_{\chi_{1}}(X_{1})\cdots \delta J_{\chi_{N}}(X_{N})}\right |_{J=0}\ .
\label{Green}
\ee
From the definition of $Z[J]$, we deduce the canonical dimensions and
helicities of the source fields, displayed in Table 2~: 
$$
\stackrel{Table\;2\;:\;Sources}{
\begin{tabular}{|c|c|c|c|c|}
\hline 
& $ $ & $ $ & $ $ & $ $  \\
& $J_{u}$&$J$ & $\bar{J} $ & $J_{b}$  \\ \hline
$\dim $ & $2$&$2$ & $2$ & $1$  \\ \hline
$\mathrm{hel}$ & $0$&$1$ & $-1$ & $0$  \\ \hline 
\end{tabular}
}  
$$
We can now derive from (\ref{Z}) the equations of  motion
\begin{eqnarray}
ik\left(\bar{\partial}A_{u}-\partial_{u}\bar{A}\right)+\bar{J}&=&0
\nonumber \\
ik\left(\partial_{u}A-\partial A_{u}\right)+J&=&0\label{eq_motion}\\
ik\left(\partial \bar{A}-\bar{\partial}A+\f{1}{k}b
\right)+J_{u}&=&0\nonumber\\
iA_{u}+J_{b}&=&0\nonumber
\end{eqnarray}
which lead to the local Ward identity
\be
\partial \bar{J}+\bar{\partial}J +\partial_{u}
J_{u}+i\partial_{u}b=0\ .
\label{Ward}
\ee
Notice that, it is because of the axial gauge condition \equ{gc} that 
the gauge symmetry of the model is described by a local Ward identity.

\section{Introduction Of A Boundary}

Let us now introduce as a boundary the plane 
\be
u=0\ .
\label{boundary}
\ee

Following
\cite{Symanzik81,Blasi91, Piguet91}, the presence of the boundary is 
correctly taken into account if the following two constraints are 
satisfied~:
\begin{enumerate}
\item Separability: every Green
function (\ref{Green}), in particular the propagators, 
must vanish when the points
$X_{n}$ don't lie in the same half-space delimited by the
boundary \equ{boundary}.
\item Locality: if all the points $X_{n}$ lie in the same
half-space, and none of them is on the boundary, the
correlators are the same as in the original theory without
boundary.
\end{enumerate}

The two conditions are satisfied by a generating functional of the 
form
\be
Z=Z_{+}+Z_{-}\ ,
\ee
where the indices + and - refer to the half-spaces $u>0$ and $u<0$, 
respectively.

For what concerns the propagators, the above constraints
are satisfied if
\be
\Delta_{\chi_{1}\chi_{2}}(X_{1},X_{2})=
\vev{\chi_{1}(X_{1})\chi_{2}(X_{2})}=
\theta_{+}\Delta_{+}(X_{1},X_{2})+\theta_{-}\Delta_{-}(X_{1},X_{2})
\ee
where
\be
\theta_{\pm}\equiv\theta(\pm u_{1})\theta(\pm u_{2})
\ee
is equal to $1$ if $u_{1}$ and $u_{2}$ are both in the same half-space, and
equal to zero otherwise.\newline
The elements of the matrix $\Delta_{\chi_{1}\chi_{2}}(X_{1},X_{2})$ 
have been determined in \cite{Piguet91}, and the aim of this paper is 
to find out which quantum fields can describe the physics on the 
two-sided two dimensional (2D) boundary, and by means of which 2D 
action. To reach this goal, we derive
the correlation functions on each side of the boundary, starting from
the 2D Ward identity expressing the gauge invariance on the boundary, 
and after that we will look for an
effective action which reproduces these correlation
functions.\newline 
We need to know how the equations of motion
(\ref{eq_motion}) and the Ward identity \equ{Ward} are modified by 
the presence of the boundary.\newline
Taking into account power counting and helicity constraints, the 
presence of the boundary modifies the equations of motion 
\equ{eq_motion} as 
follows \cite{Piguet91}~:
\begin{eqnarray}
ik
\left(\bar{\partial}A_{u}-\partial_{u}\bar{A}\right)+\bar{J}&=&ik\delta(u)
\bar{A}_{-}\nonumber\\
ik\left(\partial_{u}A-\partial A_{u}\right)+J&=&ik\delta(u) A_{+}
\label{eq_motion_boundary}\\
ik\left(\partial \bar{A}-\bar{\partial}A
+\frac{1}{k}b\right)+J_{u}&=&0\nonumber\\
i A_{u}+J_{b}&=&0\nonumber\ ,
\end{eqnarray}
where
\be
\bar{A}_{\pm}(Z) \equiv \lim_{u\to 0^{\pm}} 
    \bar{A}(X) \, \, \, \,
\,\,\,\,\,\,\,\, 
    A_{\pm}(Z) \equiv \lim_{u\to 0^{\pm}} 
    A(X)
\label{boundaryfields}\ee 
(with $(Z)\equiv (z,\bar{z})$),
are the Chern-Simons fields on the boundary, 
and, as it has been shown in \cite{Blasi91,Piguet91}, 
the following 
Dirichlet boundary conditions must hold
\be
\bar{A}_{+}(Z) = A_{-}(Z) =0\ .
\label{bc}\ee
Correspondingly, the local Ward identity \equ{Ward} acquires a
boundary 
breaking~:
\be
\partial \bar{J}+\bar{\partial}J +\partial_{u}
J_{u}+i\partial_{u}b=ik \delta(u)\left(\bar{\partial}A_{+}+\partial
\bar{A}_{-}\right)\ ,
\label{Ward_boundary}
\ee 
which, once integrated, gives the 2D Ward identity describing the 
gauge symmetry on the boundary
\be
-\f{i}{k}\int^{+\infty}_{-\infty}du\left[\partial
\bar{J}(X)+\bar{\partial}
J(X)\right]=\bar{\partial}A_{+}(Z)+\partial
\bar{A}_{-}(Z)\ .
\label{Ward_integrated}
\ee
We concentrate ourselves on the side 
$u\rightarrow 0^{+}$, the expressions for the opposite side
being derived by means of the
inversion \equ{inversion}.\newline
We begin by the correlator
\be
\vev{A_{+}(Z)A_{+}(Z')}=
\left. 
\lim_{u,u'\rightarrow 0^{+}}
\f{\delta^{2} Z}{\delta \bar{J}(X)\delta \bar{J}(X')}
\right |_{J=0}\ .
\label{aa}
\ee
Applying to both sides of \equ{Ward_integrated} 
the functional operator
\be
\lim_{u\rightarrow 0^{+}}\f{\delta}{\delta \bar{J}(X)}\ ,
\ee
we obtain the differential equation between correlators
\be
-\f{i}{k}\partial\delta^{2}(Z-Z') =
\bar\partial\vev{A_{+}(Z)A_{+}(Z')}
+ \partial\vev{\bar{A}_{-}(Z)A_{+}(Z')}
\ .
\ee
Remembering the separability condition according to which 
correlation functions between points belonging to different 
half-spaces vanish, using the helicity and power counting
constraints, 
and, finally,  exploiting  the relation
\be
\delta^{2}(Z-Z')=\f{1}{2\pi i} \bar{\partial}
\f{1}{z-z'}
\ee
between tempered distributions, one has
\be
\vev{A_{+}(Z)A_{+}(Z')}=\f{1}{2\pi k} \f{1}{(z-z')^{2}}\ .
\ee
Notice that the above correlator turns out to be chiral.
Proceeding in an analogous way, from \equ{Ward_integrated} we
obtain also
\be
-\f{i}{k}\bar\partial\delta^{2}(Z-Z') =
\bar\partial\vev{A_{+}(Z)\bar{A}_{+}(Z')}
+ \partial\vev{\bar{A}_{-}(Z)\bar{A}_{+}(Z')}
\ ,
\ee
which gives
\be
\vev{A_{+}(Z)\bar{A}_{+}(Z')} = 
-\f{i}{k}
\delta^{2}(Z-Z')\ .
\ee
For what concerns the propagator 
$\vev{\bar{A}_{+}(Z)\bar{A}_{+}(Z')}$, it is set to zero, since it is 
not generated by the Ward identity \equ{Ward_integrated}~:
\be
\vev{\bar{A}_{+}(Z)\bar{A}_{+}(Z')}=0\ .
\ee
The following boundary correlator matrix summarizes our results~:
\be
\left(
\begin{array}{ccc}
\vev{A_{+}(Z)A_{+}(Z')}
&\vev{A_{+}(Z)\bar{A}_{+}(Z')} \\
\vev{\bar{A}_{+}(Z)A_{+}(Z')} &
\vev{\bar{A}_{+}(Z)\bar{A}_{+}(Z')}
\end{array}\right)=\left(
\begin{array}{ccc}
\f{1}{2\pi k} \f{1}{(z-z')^{2}}&\!\!
-\f{i}{k}\delta^{2}(Z-Z')\\
-\f{i}{k}\delta^{2}(Z-Z')&0
\end{array}\right)
\label{matrix}
\ee
We are left now with the task of finding an effective 2D theory on 
the boundary, which reproduces the correlation functions 
\equ{matrix}.

\section{Effective Bosonic Action}

Let us consider now the 2D action $S_{B}$, living on the external 
side of the boundary $u=0^{+}$~:
\be
S^{(+)}_{B}=i \f{k}{2} \int dz d\bar{z} \left[\partial \varphi_{+}
\bar{\partial}\varphi_{+} 
+\left(\bar{\partial}\varphi_{+}\right)^{2}\right]\ ,
\label{bosonic_action}
\ee
where $\varphi_{+}(Z)$ is a bosonic field whose propagator is (see 
Appendix \ref{appendix})~: 
\be
G(z-z')=\vev{\varphi_{+}(Z)\varphi_{+}(Z')}=\f{1}{2\pi k}\ln
\frac{(z-z')}{\mu}.
\label{propagator_bosonic}
\ee
From \equ{propagator_bosonic}, the following correlators 
are easily derived~:
\bea
\vev{\partial \varphi_{+}(Z)\partial\varphi_{+}(Z')}&=&\f{1}{2\pi
k}\f{1}{(z-z')^{2}}\\
\vev{\partial\varphi_{+}(Z)\bar{\partial}\varphi_{+}(Z')}&=&
-\f{i}{k}\delta^{2}(Z-Z')\\
\vev{\bar {\partial}\varphi_{+}(Z)\bar{\partial}\varphi_{+}(Z')}&=&0\ 
,
\eea
which coincide with those appearing in \equ{matrix}. We are thus led 
to identify the components of the gauge fields on the boundary 
$u=0^{+}$ with the
derivative of a bosonic field, through the relations
\begin{eqnarray}
A_{+}(Z)&\leftrightarrow &\partial \varphi_{+}(Z)\\
\bar{A}_{+}(Z)&\leftrightarrow &\bar{\partial}\varphi_{+}(Z)\ .
\end{eqnarray}
Coming back to the original, euclidean, coordinates 
$(x_{0},x_{1},x_{2})$, the chiral action
(\ref{bosonic_action}) reads
\be
S^{(+)}_{B}=-\f{k}{2}\int dx_{1} dx_{0}
\left[\left(\partial_{1}\varphi_{+}\right)
\left(v\partial_{1}\varphi_{+}-i\partial_{0}\varphi_{+}\right)\right]\
,
\label{chiralaction}
\ee
whose equation of motion is
\be
\left(v \partial_{1}-i\partial_{0}\right)\partial_{1}\varphi_{+}=0
\label{eom}\ee
which implies the chirality of $\partial_{1}\varphi_{+}$. 
In this case
the direction of the propagation for the field is regressive.\newline
We stress that the equation of motion \equ{eom} alone does not 
imply chirality also for the undifferentiated bosonic field 
$\varphi_{+}(Z)$. In order to be able to claim that the bosonic field 
is indeed a chiral function of the coordinate $z$ only, some 
additional boundary conditions must be invoked 
\cite{Cappelli95,Sonn88}.
For what concerns the internal side of the boundary, it is 
straightforward to obtain another bosonic action, related 
to (\ref{bosonic_action}) by
the inversion \equ{inversion}, depending on a 
bosonic field $\varphi_{-}(Z)$, whose equation of motion
implies a progressive propagation for 
$\partial_{1}\varphi_{-}$. \newline
From the expression (\ref{chiralaction}), we have the hamiltonian 
\be
H=v\f{k}{2}\int dx_{1} \left(\partial_{1}\varphi_{+}\right)^{2}\ ,
\label{ham}\ee
which, coincides with the one proposed 
by Wen \cite{Wen90} to describe
the edge states in the FQHE for the
Laughlin's sequence by imposing
\be
k=\f{1}{2\pi \nu}
\label{kfilling}
\ee
as in equation (\ref{nu}).\newline
Therefore, an abelian Chern-Simons
gauge field theory with boundary admits, on the opposite sides of the
boundary, a description in term of a chiral bosonic field whose
propagation is regressive for the external side of the boundary and
progressive for the internal side.\newline
The importance of our result is that we derived it only by using the
Symanzik's separability condition and by the identification of the
correlators of the gauge field with the second derivative of the
propagator of a scalar field living on the boundary. We don't impose
any other additional condition on the boundary.\newline

\section{Conserved Boundary Currents} \label{Chapter5}

Let us consider the following operators defined on the external side
of the boundary $u\rightarrow 0^{+}$
\bea
K_{+}(Z)&\equiv&\f{1}{2\pi}\lim_{u\rightarrow 0^{+}}A(X)\label{k}\\
\bar{K}_{+}(Z)&\equiv&\f{1}{2\pi}\lim_{u\rightarrow 0^{+}}\bar{A}(X)\
,
\label{bk}\eea
while the analogous expressions $K_{-}(Z)$ and
$\bar{K}_{-}(Z)$, as usual, can be derived by the 
inversion \equ{inversion}.\newline
From \equ{bc} and \equ{Ward_integrated}, we have 
\cite{Blasi91,Piguet91}
\bea
\bar{\partial}K_{+}(Z)&=&0
\label{condition1}\\
\bar{K}_{+}(Z)&=&0\ .
\label{condition2}
\eea
The operators \equ{k} and \equ{bk} satisfy the following commutation 
relation 
\be
[K_{+}(z),K_{+}(z')]=-\f{i}{(2\pi)^{2}k}\partial\delta(z-z')\ ,
\label{km}\ee
as it can be immediately derived from the Ward identity 
\equ{Ward_integrated}. This commutation relation is the remnant of 
the Kac-Moody algebra formed by the nonabelian counterpart of the 
chiral conserved currents $K_{+}(z)$ \cite{Blasi91,Piguet91}. Under 
this respect, the coefficient
$\displaystyle{-\f{i}{(2\pi)^{2}k}}$ can be seen as the central
charge 
of the Kac-Moody algebra.\newline
Moreover, \equ{condition1} and \equ{condition2} obviously yield 
\be
\bar{\partial}K_{+}+\partial \bar{K}_{+}=0\ ,
\label{conservation1}
\ee
which is easily interpreted as a conservation relation.
In fact, coming back
to the euclidean coordinates, we can write $K_{+}$ and $\bar{K}_{+}$
in terms of new quantities $\rho_{+}$ and $J_{+}$,  respectively
time and space components of a vector in the euclidean spacetime, in 
analogy with \equ{lcfields}~:
\bea
K_{+}&\equiv&\f{1}{\sqrt{2}v}\left(J_{+}+iv\rho_{+}\right)\\
\bar{K}_{+}&\equiv&\f{1}{\sqrt{2}v}\left(J_{+}-iv\rho_{+}\right)\ .
\eea
In terms of  $\rho_{+}$ and $J_{+}$, the relation \equ{conservation1}
becomes
\be
\partial_{1}J_{+}+\partial_{0}\rho_{+}=0\ ,
\ee
which is the familiar continuity equation involving a density
$\rho_{+}$ and a 
current $J_{+}$.

From (\ref{condition1}) and (\ref{condition2}), we also
get the chirality of $\rho_{+}$ 
\be
v\partial_{1}\rho_{+}-i\partial_{0}\rho_{+}=0\label{5.9}\\
\ee
and the identification of $\rho_{+}$ and $J_{+}$~:
\be
J_{+} = iv\rho_{+}\ .
\ee
On the other hand, the commutation relation \equ{km}, in terms of the 
density $\rho_{+}(z)$ is
\be
[\rho_{+}(z),\rho_{+}(z')] =
i\frac{\nu}{4\pi}\partial\delta(z-z')\ ,
\label{kmdensity}
\ee
where we introduced the 
filling factor $\nu$ by means of \equ{kfilling}. The commutation 
relation \equ{kmdensity}, derived here in the framework of a gauge 
field theory with boundary, quite remarkably turns out to coincide
with the relation 
peculiar of the Tomonaga-Luttinger theory 
\cite{tomolutti,luttinger,haldane} for a 1+1 
dimensional liquid of interacting electrons.\newline
Finally, comparing \equ{5.9} with the equation of motion \equ{eom},
we can make the identification between real hermitian operators
\be
\rho_{+} = -\f{i}{2\pi}\partial_{1}\varphi_{+}\ .
\label{rhophi}
\ee 
Coming back to the Minkowskian spacetime, the electronic density reads
\be
\rho^{M}_{+} = -\f{1}{2\pi}\partial_{1}\varphi_{+}\ .
\label{rhophimink}
\ee

In the usual model of the edge states of the FQHE \cite{Wen90},
$\rho^{M}_{+}$ corresponds to the electron density on the edge the Hall
bar and the above relation allows to connect this physical quantity
to the chiral bosonic field that propagate along the edge.

\section{Conclusions}

In this paper we have shown how the well known results concerning the 
edge states
in the FQHE can be deduced by means of Symanzik's general approach to 
Quantum Field Theories with boundary.
In particular, we derived the bosonic chiral hamiltonian written 
in terms of an electronic density, 
and the Luttinger-Tomonaga commutation relations for an
electronic liquid in 1+1 dimensions. \newline
The main   point  we tried to clarify, is that there is a   method 
which respects all the basic assumptions of perturbative
quantum field theory, by which all the rest can be deduced.
This not only reinforces the validity of the known results, but 
also gives us hope of going further and apply the same method 
to describe more complex sequences of the FQHE as the Jain 
sequence \cite{Jain89} or non abelian states.\newline
Work is in progress in these directions \cite{progress}.

%
%
%
%
\appendix

\section{Bosonic Chiral Propagator}\label{appendix}
In this appendix we want to derive the expression
(\ref{propagator_bosonic}) for the correlator of the bosonic
field.\newline
From the action (\ref{bosonic_action}) we obtain the differential
equation for the Green function $G(Z)$

\begin{equation}
-k (-i\partial _0 +v \partial  _1 )\partial _1 G(Z) = \frac{v}{i}
\delta^2 (Z)\ .
\end{equation}

Introducing the function
\be
f(Z) = \partial _1 G(Z)
\ee
that satisfies the relation
\begin{equation}
\bar{\partial} f(Z) =  \frac{i}{\sqrt{2}k}\delta ^2 (Z)\ ,
\label{equation_f}
\end{equation}
we find the solution of (\ref{equation_f}) 
\be
f(z)=\frac{1}{2\sqrt{2}\pi k}\frac{1}{z}\ .
\ee
Notice that $f$ is a chiral function in agreement with the result in
Section \ref{Chapter5}.\newline
Finally we can integrate over $x_{1}$ to obtain 
\be
G(z) = \frac{1}{2\pi k}\ln(\frac{z}{\mu} )
\ee
where we introduce the arbitrary scale $\mu$. In this work we
identify the derivatives of $G(z)$ with the correlators of the
Chern-Simons abelian field on the boundary, therefore this parameter
doesn't appear in these physical observables.


\begin{thebibliography}{999}
\bibitem{Wen90}     X. G. Wen, \emph{Phys. Rev. B} \textbf{41}, 12838 (1990).
\bibitem{Wen91}     X. G. Wen, \emph{Int. J. Mod. Phys. B} \textbf{6},
                    1711 (1992).
\bibitem{Prange}    R. E. Prange and S. M. Girvin, eds, 
                    ``{\it The Quantum Hall Effect }'', 
		    (Springer-Verlag, 2nd ed. 1990).
\bibitem{Halperin}  B. I. Halperin, \emph{Phys. Rev. B} \textbf{25}, 2185 (1982).
\bibitem{Cappelli95}A. Cappelli, G.V. Dunne, C. Trugenberger, G. Zemba, 
                    \emph{Nucl. Phys.} \textbf{B398}, 531 (1993).
\bibitem{Kane95}    C. L. Kane, M. P. A. Fisher, \emph{Phys. Rev. B}
                    \textbf{51}, 13449 (1995).
\bibitem{Fradkin99} E. Fradkin, A. Lopez, \emph{Phys. Rev. B}
                    \textbf{59}, 15323 (1999).
\bibitem{Symanzik81}K. Symanzik, \emph{Nucl. Phys.} \textbf{B190}, 1
                    (1981).
\bibitem{Blasi91}   A. Blasi, R. Collina, \emph{Int. J. Mod. Phys. A}
                    \textbf{7}, 3083 (1992).
\bibitem{Piguet91}  S. Emery, O. Piguet, \emph{Helv. Phys. Acta}
                    \textbf{64}, 1256 (1991).
\bibitem{Laughlin83}R. B. Laughlin, \emph{Phys. Rev. Lett.} \textbf{50}, 
                    1395 (1983).
\bibitem{Sonn88}    J. Sonnenschein, \emph{Nucl. Phys.} \textbf{B309}, 
                    752 (1988).
\bibitem{tomolutti} S. Tomonaga, \emph{Progr. Theor. Phys. (Kyoto)}
                    \textbf{5}, 544 (1950).
\bibitem{luttinger} J. M. Luttinger, {\it J. Math. Phys.}
                    \textbf{4}, 1154 (1963).
\bibitem{haldane}   F. D. M. Haldane, \emph{J. Phys. C}
                    \textbf{14}, 2585 (1981).
\bibitem{Jain89}    J. K. Jain, \emph{Phys. Rev. Lett.} \textbf{63}, 199 (1989).
\bibitem{progress}  A. Blasi, D. Ferraro, N. Maggiore, N. Magnoli, M. 
                    Sassetti, \emph{work in progress}.
\end{thebibliography}
\end{document}